\documentstyle[12pt]{article}
\title{Quantum black-hole information missing in the semiclassical treatment}
\author{Hrvoje Nikoli\'c \\
Theoretical Physics Division, Rudjer Bo\v{s}kovi\'{c} Institute, \\
P.O.B. 180, HR-10002 Zagreb, Croatia \\
{\normalsize hrvoje@thphys.irb.hr} \\
\makebox[1in]{} \\
}
\date{\today}
\begin{document}
\maketitle
\begin{abstract}
In the semiclassical treatment of gravity, an external observer can measure
only the mean (not the exact) mass of the black hole (BH).
By contrast, in fully quantum gravity the exact (not only mean) 
BH mass is measurable by the external observer.
This additional information (missing in the semiclassical treatment) 
available to the external observer significantly helps to understand how
information leaks out during the BH evaporation. 
\end{abstract}
\vspace*{0.5cm}

\section{Introduction}

According to the semiclassical theory of gravity, in which matter is quantized
while gravity is treated classically, the black-hole (BH) evaporation
\cite{hawk1} leads to a transition from a pure to a mixed state \cite{hawk2},
contradicting unitarity of quantum mechanics. Now most physicists, 
including Hawking \cite{hawk3}, agree that the true process of BH evaporation
in fully quantum gravity is unitary, but it is still far from clear how exactly unitarity restores
and what exactly is wrong with the semiclassical treatment.
(For reviews see, e.g., \cite{har, pag, gid, str}.)
There is a lot of evidence, especially
from string theory \cite{peet}, suggesting that black holes evaporate completely
in a unitary manner, which
implies that all BH information must leak out with the radiation. Nevertheless, such a leak of information seems to contradict the principle of locality, while
the true nature of such nonlocality is still a controversial issue. 

In this paper we attempt to provide a better understanding of the process 
of unitary BH evaporation. In particular, we want to explain how information
leaks out from a fully quantum black hole and to identify the physical origin 
of nonlocality associated with it. The power (as well as the weakness) of
our approach lies in the fact that we only use the general
principles of quantum mechanics (such as the basic facts on
quantum entanglement and quantum information), without specifying any details 
specific to (still purely understood) quantum gravity, except some  
classical properties of gravity
that are expected to be valid on the quantum level as well.

In the next section we start by explaining the conceptual difference between 
the semiclassical BH mass and the quantum BH mass. The quantum uncertainty
of radiation energy and BH mass is discussed quantitatively in Sec.~\ref{SEC3}.
The central section is Sec.~\ref{SEC4}, where we explain how quantum
measurements extract information from black holes and introduce a 
non-unitary aspect of BH evaporation. Sec.~\ref{SEC5} discusses
nonlocal aspects as well as other
fundamental interpretational issues associated with quantum measurements. 
A summary of our results is given in Sec.~\ref{SEC6}.

\section{Black-hole mass: semiclassical vs quantum}
\label{SEC2}

In semiclassical gravity one assumes that the internal quantum BH degrees 
of freedom are not available to an external observer. Consequently, 
the information available to the external observer 
is obtained by tracing out over the internal degrees of freedom, leading
to the external density matrix of Hawking radiation \cite{bd}
\begin{equation}\label{rho}
\rho=\prod_{\omega}\sum_{n} |f_{\omega,n}|^2 
|n,\omega\rangle \langle n,\omega| ,
\end{equation}
which is a mixed thermal state.
Here $f_{\omega,n}=\sqrt{1-e^{-\beta\omega}} e^{-\beta\omega n/2}$,
$|n,\omega\rangle$ is the state containing $n$ particles
with the energy $\omega$, and $\beta=8\pi M$ is the inverse temperature
of the black hole with the mass $M$.  

The crucial assumption leading to (\ref{rho}) is that the external
observer knows nothing about the internal degrees of freedom.
However, this assumption is {\em not} really true! According to the
classical no-hair theorem, an external observer knows {\em almost} nothing
about the interior, {\em except} the interior mass $M$ (as well as the charge and 
the angular
momentum which, for simplicity, are assumed to be zero). The mass is 
measured by an external observer through a measurement of the external gravitational field. Is that additional information missing in (\ref{rho})?
At first sight this information is already there, because (\ref{rho}) explicitly
depends on $M$ through the inverse temperature $\beta$. However, 
the parameter $M$ in (\ref{rho}) is a {\em classical} parameter, not a
quantum one. On the other hand, in a fully quantized theory of gravity,
the mass also must be a quantum observable. As we shall see, a 
quantum treatment of the BH mass measurable by an external observer
significantly changes $\rho$ describing the information available
to the external observer.

In the world in which both gravity and matter obey quantum laws, 
what exactly does it mean that the mass $M$ is ``classical"?
This means that it is actually the {\em mean} value 
$\langle\Psi| \hat{M} |\Psi\rangle$ of the mass, 
calculated with respect to a fully quantum state $|\Psi\rangle$. Indeed, 
in semiclassical gravity, the gravitational field 
is a classical field determined by the mean energy-momentum
of quantum matter. Nevertheless, although in some cases
such a semiclassical treatment may serve as a good
approximation, there is an experimental proof \cite{page} 
that {\em the measured gravitational
force is not really given by the mean energy-momentum, but by the
actual energy-momentum which in some cases may significantly
deviate from the mean value}. For example,
for a superposition $|\Psi\rangle = (|M_1\rangle + |M_2\rangle)/\sqrt{2}$
of the mass eigenstates $|M_1\rangle$ and $|M_2\rangle$, the mean value
is $(M_1+M_2)/2$, while the measured mass is either $M_1$ or $M_2$.
Therefore, a correct
quantum treatment of BH evaporation should describe the BH mass
as a genuine quantum observable. What we find below is that the
essential consequences of such a treatment can be understood without
knowing the details of otherwise not so well understood  
theory of quantum gravity.

\section{Uncertainty of energy}
\label{SEC3}

The state (\ref{rho}) involves product states of the form
$|n_1,\omega_1\rangle |n_2,\omega_2\rangle \cdots $, the energy of which is
equal to
${\cal E}=n_1\omega_1+n_2\omega_2+\ldots$. Various states of that form
are involved in (\ref{rho}), so
the actual energy ${\cal E}$ in the state (\ref{rho}) is uncertain. Only the mean energy of (\ref{rho}) is well
defined. To make this uncertainty of energy more explicit, it is convenient to
work in a basis consisting of energy eigenstates $|{\cal E},i\rangle$,
where $i$ labels all orthogonal states having the same total energy ${\cal E}$.
(In particular, different values of $i$ include different total
numbers of particles.)
This basis is a basis for the external degrees of freedom. Similarly,
the internal BH degrees of freedom are spaned by the basis $|M,I>$, where
$M$ is the BH mass, while $I$ labels orthogonal states having the same $M$.
Let the initial BH mass (i.e., the mass before the process of 
Hawking radiation has started) be $M_0$.
Then the most general pure state consistent with energy conservation is
\cite{niksc} 
\begin{equation}\label{Psi}
|\Psi\rangle =\sum_{\cal E} \sum_I \sum_i D_{{\cal E},I,i} |M_0-{\cal E},I> \otimes |{\cal E},i\rangle .
\end{equation}
In accordance with the general principles of quantum mechanics, energy is 
conserved exactly, not only in average, in the sense that 
(\ref{Psi}) is a superposition of states each of which has the same total
energy $M+{\cal E}=(M_0-{\cal E})+{\cal E}=M_0$ equal to the initial energy $M_0$.
The coefficients $D_{{\cal E},I,i}$ depend on details of the dynamics described
by quantum gravity. In the case of BH radiation, these coefficients are 
time dependent, while energy ${\cal E}$ represents the total energy
of all particles radiated at different times and temperatures \cite{nikbhinf}. 
By tracing out over the internal degrees 
of freedom, (\ref{Psi}) leads to the external density matrix
\begin{equation}\label{rho2}
\rho=\sum_{\cal E} \tilde{\rho}_{\cal E} ,
\end{equation}
where
\begin{equation}
\tilde{\rho}_{\cal E} \equiv \sum_I \sum_{i,i'} D_{{\cal E},I,i} 
D^*_{{\cal E},I,i'}
|{\cal E},i\rangle \langle {\cal E},i'| .
\end{equation}
The tilde above $\tilde{\rho}_{\cal E}$ denotes that the trace of this 
density matrix is not normalized to unity. The corresponding normalized
density matrix is 
\begin{equation}
\rho_{\cal E}=\frac{\tilde{\rho}_{\cal E}}{{\rm Tr}\tilde{\rho}_{\cal E}} .
\end{equation}
During a short time after the start of radiation (during which the change
of the BH temperature can be neglected), the time dependence of the 
coefficients $D_{{\cal E},I,i}$ can be neglected. Furthermore,
during this short time the coefficients $D_{{\cal E},I,i}$ may be such that
the density (\ref{rho2}) coincides with (\ref{rho}) \cite{nikbhinf}.
Indeed, both (\ref{rho}) and (\ref{rho2}) have uncertain radiation
energy ${\cal E}$. 

\section{Extraction of information by quantum measurement}
\label{SEC4}

Now the crucial observation is the fact that the state (\ref{rho2}) 
does {\em not} represent all
information available to the external observer. Instead, as we have
explained in Sec.~\ref{SEC2}, the external observer
can also measure the exact BH mass $M$. From (\ref{Psi}) we see
that the measurement of mass also determines the exact radiation energy
${\cal E}=M_0-M$. 
Thus, {\em the measurement induces the state reduction}
\begin{equation} 
\rho\rightarrow\rho_{\cal E} .
\end{equation}
Therefore, the information available to the external observer
is not given by the whole sum (\ref{rho2}), but only by one of the
terms in (\ref{rho2}), namely by $\rho_{\cal E}$ which contains
more information than $\rho$. In particular, during a short time after
the start of radiation as above, $\rho_{\cal E}$ 
corresponds to the thermal state in the grand {\em micro}canonical ensemble,
which should be contrasted with (\ref{rho}) that corresponds
to the thermal state in the grand canonical ensemble \cite{nikbhinf}. 

For a given energy of radiation ${\cal E}$ and the BH mass $M=M_0-{\cal E}$, the external
observer still does not possess information on the internal BH degrees of freedom
labeled by $I$. This is why $\rho_{\cal E}$ is still a mixed state. However, 
it is reasonable to assume that $I$ attains a smaller number of different
values for smaller $M$, i.e., that a lighter black hole 
contains a smaller number of internal degrees of freedom. Indeed, the results
from string theory confirm that assumption \cite{peet,hor}.
Therefore, {\em as the total radiated energy ${\cal E}$ increases with time,  
the state $\rho_{\cal E}$ becomes less mixed} (even if
the unreduced state $\rho$ becomes more mixed with time). In particular, if the 
black hole eventually evaporates completely, and if the final massless
``black hole" does not contain any internal degrees of freedom
(i.e., $|0,I>\equiv|0>$), then the final state of radiation is 
\begin{equation}\label{e1}
\tilde{\rho}_{M_0}  =  \sum_{i,i'} D_{M_0,i} D^*_{M_0,i'}
 |M_0,i\rangle \langle M_0,i'| , 
\end{equation}
where $|M_0,i\rangle$ are states of radiation the total energy of which
is equal to the initial BH mass $M_0$. Since (\ref{e1}) 
does not involve summation
over the internal degrees of freedom labeled by $I$,
this final state is actually a {\em pure} (no longer mixed) state
\begin{equation}
\tilde{\rho}_{M_0}  =  |\tilde{M_0}\rangle \langle \tilde{M_0}| , 
\end{equation}
where
\begin{equation}\label{e3}
|\tilde{M_0}\rangle \equiv \sum_i D_{M_0,i} |M_0,i\rangle .
\end{equation}
This is how complete BH evaporation becomes compatible with unitarity;
{\em unitarity refers to the unreduced state (\ref{Psi}) or (\ref{rho2}),
while the measurable process of evaporation is associated with the reduced state
$\rho_{\cal E}$}. (It is analogous to the fact that a measurable cat is a reduced either dead or alive cat, even though the unitary Schr\"odinger cat involves a 
superposition of a dead and alive cat.)

Of course, our analysis does not imply that the black hole must necessarily
evaporate completely. The ultimate fate of the quantum black hole depends on details of the correct theory of quantum gravity. Our equations
(\ref{e1})-(\ref{e3}) correspond to the {\em assumption} that black holes
evaporate completely, which, however, may be a wrong assumption.
As discussed in \cite{nikbhe1} and \cite{nikbhinf}, 
one of the crucial questions that quantum gravity needs to answer 
is whether the degrees of freedom carried by the negative-energy particles absorbed by the black hole are physical or not. 
While they are physical
according to the semiclassical treatment of BH radiation \cite{nikbhe1}, string theory suggests that they are unphysical \cite{nikbhinf}. 
Nevertheless, independently on these unresolved issues, our result
that the information available by the external observer is given by
$\rho_{\cal E}$ rather than $\rho$, which implies that this information may be
significantly larger than that in the semiclassical treatment, remains intact.

\section{The role of entanglement nonlocality} 
\label{SEC5}

As we have seen, information leaks out from the black hole by measuring
the BH mass as a quantum observable. To acquire a better understanding of the physical mechanism that lies behind such a measurement, assume that the 
only quantity {\em directly} measured by an external observer is the
energy of radiation ${\cal E}$. From (\ref{Psi}) we see that the energy of radiation
is perfectly correlated with the mass of the black hole, such that their
sum is allways equal to $M_0$. Thus, the direct measurement of radiation energy
may be thought of as an indirect measurement of the BH mass. 
(Indeed, an external observer
could, in principle, jump into the black hole to directly verify
if his indirect measurement of the BH mass was correct. According to the standard rules of quantum mechanics, he would obtain that it was.)
But how an external observable such as ${\cal E}$ can be perfectly correlated with an
internal observable such as $M$? Does it contradict the classical principle
of locality? It certainly does. Nevertheless, it is in a perfect
accordance with the already known principles of physics.
Namely, the nonlocal nature of quantum correlations caused by quantum
entanglement is a well understood (see, e.g., \cite{bell})
and experimentally tested fact. 
Thus, the nonlocal mechanism that
allows information to leak out from the black hole is nothing but a special
case of the well-known quantum effect, 
the effect known also under the name Einstein-Podolsky-Rosen \cite{epr} correlations. 

Note also that the results above do not depend on the controversial issues
related to the interpretation of quantum mechanics. Nevertheless, 
some specific interpretations may further clarify the physical mechanism
responsible for such nonlocal correlations. For example, in the 
Copenhagen interpretation the measurement induces an instantaneous nonlocal collapse of the wave function. Thus, the measurement of the BH mass
is thought of as a collapse $\rho\rightarrow\rho_{\cal E}$.
One of the problems with the collapse 
is an explicit violation of unitarity. An alternative, especially
popular among some quantum cosmologists, is the many-world interpretation. 
This interpretation avoids the need for a wave-function collapse
and makes quantum mechanics look more local (see, however,
\cite{niknonloc}). 
Yet another interpretation that avoids the collapse is the Bohmian
interpretation, in which the actual objective physical reality
corresponds to a nonlocal ``hidden variable''
not identified with the quantum state.
Unlike the many-world interpretation,
the Bohmian interpretation describes only one explicitly nonlocal world. 
Some recent results
indicate that the Bohmian interpretation could be intimately related to
the general-covariant quantization of fields \cite{nikbg1,nikbg2} and string theory \cite{niks1,niks2,niks3,niks4}, which makes it promissing for the application to quantum physics of black holes.

\section{Summary}
\label{SEC6}

Our results can be summarized by the statement that the 
so-called black-hole information
paradox can be viewed as a version of the Schr\"odinger cat paradox.
Namely, by going beyond the usual semiclassical description one necessarily
treats the BH mass as a quantum observable, implying that any observation
of the BH mass induces a state reduction that seemingly violates unitarity.
Hence, the observed black hole is not the same thing as the unitary evolving
quantum BH state, which is a fact that seems to be widely unrecognized in 
the existing discussions of the BH information paradox.
In particular, it is possible that the observed black hole has completely evaporated,
but that, at the same time, the unitary evolving BH state involves a superposition
of evaporated and non-evaporated black holes with a mean mass larger than zero.
In this sense, the BH evolution can be viewed as 
being both nonunitary and unitary, 
depending on whether the effects of quantum measurements 
are taken or not taken into account, respectively.
We have also explained how quantum entanglement accounts for
nonlocality needed to make the complete evaporation consistent
with unitarity. 

\section*{Acknowledgements}

This work was supported by the Ministry of Science of the
Republic of Croatia under Contract No.~098-0982930-2864.

\end{document}